\documentclass[ALICE,manyauthors]{cernphprep}

\usepackage{lineno}

\usepackage[usenames,dvipsnames]{xcolor}
\usepackage[bookmarks]{hyperref}

\newcommand{\Om}{$\Omega^-$}
\newcommand{\Mo}{$\overline{\Omega}^+$}
\newcommand{\X}{$\Xi^-$}
\newcommand{\Ix}{$\overline{\Xi}^+$}
\newcommand{\Xis}{$\Xi^{\pm}$}
\newcommand{\Oms}{$\Omega^{\pm}$}
\newcommand{\pt}         {\ensuremath{p_{\mathrm{T}}}}  

\newcommand{\mTmZero}    {\ensuremath{m_{\mathrm{T}} - m_0}}
\newcommand{\meanpt}{$\langle p_\mathrm{T}\rangle$}
\newcommand{\nineH}{$\sqrt{s}~=~0.9$~TeV}
\newcommand{\seven}{$\sqrt{s}~=~7$~TeV}
\newcommand{\twoH}{$\sqrt{s}~=~0.2$~TeV}
\newcommand{\dndy}{d$N$/d$y$}
\newcommand{\pp}{pp}
\newcommand{\PbPb}{Pb--Pb}

\newcommand{\GeVc}{GeV/$c$}

\newcommand{\GeVmass}{GeV/$c^2$}

\begin{document}
%
%
\begin{titlepage}
\PHnumber{2012-080}                 
\PHdate{01 April 2012}              
%
%
\title{Multi-strange baryon production in pp collisions at 7 TeV with ALICE}
\ShortTitle{Multi-strange baryon production in \pp\ collisions at 7 TeV with ALICE}   
%
\Collaboration{ALICE Collaboration%
         \thanks{See Appendix~\ref{app:collab} for the list of collaboration members}}
\ShortAuthor{ALICE Collaboration}      
\begin{abstract}
A measurement of the multi-strange \X\ and \Om\ baryons and their antiparticles by the ALICE experiment 
at the CERN Large Hadron Collider (LHC) is presented for inelastic proton-proton collisions at a centre-of-mass 
energy of 7 TeV.   The transverse momentum (\pt) distributions  were studied at mid-rapidity ($|y| < 0.5$)  
in the range of 0.6 $<$ \pt\ $<$ 8.5 \GeVc\ for \X\ and \Ix\ baryons, and in the range of 
0.8 $<$ \pt\ $<$ 5 \GeVc\ for \Om\ and \Mo.  
Baryons and antibaryons were measured as separate particles and we find that the baryon to antibaryon ratio of both particle species is consistent with unity over 
the entire range of the measurement.
The statistical precision of the current data has allowed us to measure a difference between the 
mean \pt\ of \X\ (\Ix) and \Om\ (\Mo). 
Particle yields, mean \pt, and the spectra in the intermediate \pt\ range are not well described by 
the PYTHIA Perugia 2011 tune Monte Carlo event generator, which has been tuned to reproduce  
the early LHC data. The discrepancy is largest for \Om\ (\Mo). This PYTHIA tune approaches the 
\pt\ spectra of \X\ and \Ix\ baryons below \pt\ $<0.85$ \GeVc\ and describes the \X\ and \Ix\ spectra above \pt\ $>6.0$ \GeVc.  
We also illustrate the difference between the experimental data and model by comparing the corresponding ratios of 
(\Om+\Mo)/ (\X+\Ix) as a function of transverse mass.

\vspace*{0.5 cm}
\end{abstract}
\end{titlepage}
\setcounter{page}{2}
%


\section{Introduction}
\label{sec:Introduction}

The multi-strange baryons, \Om\ ($sss$) and \X\ ($dss$), are particularly important in high energy particle and nuclear physics due to their dominant strange quark (\mbox{$s$-quark}) content.  The initial state colliding projectiles contain no strange valence quark, therefore all particles with non-zero strangeness quantum number  are created in the course of the collision.  Moreover, the energy of the Large Hadron Collider (LHC) and its high luminosity allow for an abundant production of strange hadrons. These two factors make multi-strange baryons a valuable probe in understanding particle production mechanisms in high energy collisions.

We present a measurement of \Om\ and \Mo\ baryon transverse momentum (\pt)  spectra and yields in proton-proton (\pp) collisions at a centre-of-mass energy ($\sqrt{s}$) of 7 TeV, a measurement of \X\ and \Ix\ yields and spectra at the same energy, and a comparison of these data to a recent pp event generator, PYTHIA Perugia 2011 central tune (P2011).  
The measurements were obtained using the ALICE experiment \cite{Aamodt:2008zz} at the LHC. 

ALICE is a general purpose detector designed to study both \pp\ and \PbPb\ collisions at TeV-scale energies.  A six-layer silicon inner tracking system (ITS) and a large-volume time projection chamber (TPC) enable charged particle reconstruction with excellent momentum and spatial resolution in full azimuth down to \pt\ of 100 MeV/$c$.  

\section{Data sample and cascade reconstruction}
\label{sec:DataSampleAndCascadeReco}

Multi-strange baryons are studied in a sample of  approximately 130 million minimum bias
\seven\ \pp\ events, collected during the 2010 data taking. 
The sample is corrected for trigger inefficiencies and biases to recover a normalized sample of inelastic (INEL) events, as described in \cite{Alice:2012xsec}.
The events are selected within 10 cm of the detector's centre along
the beam direction, with vertex resolution in the transverse plane of a few hundred micrometres. 
The event vertex range is selected to maximize particle trajectory (track) reconstruction efficiency within the ITS and TPC volume. 

\X\ and \Ix\ (\Xis), as well as \Om\ and \Mo\ (\Oms) candidates are reconstructed at mid-rapidity ($|y|<0.5$) via
their characteristic weak decay topology,
\X(\Ix)~$\rightarrow\Lambda(\overline{\Lambda})+\pi^-(\pi^+$), and
\Om(\Mo)~$\rightarrow\Lambda(\overline{\Lambda}) + \mathrm{K}^-(\mathrm{K}^+)$, as described in
detail in \cite{Aamodt:2011zza}.  The branching ratios for these decay channels are 67.8~\% for \Oms\
baryons and 99.9~\% for \Xis. Charged particles, compatible with kaon, pion and proton hypotheses,
are identified using their energy loss in the TPC.  
The topology of the \Om\ and \X\ weak decay is cascade-like and consists of a V-shaped decay of the
daughter $\Lambda$ baryon ($\Lambda$ baryon hypothesis is identified as a ``V0") plus a negatively charged
track (h$^-$).  The same applies to antibaryons, however in that case the decay products are the $\overline{\Lambda}$ daughter particles and a positively charged track (h$^+$). In general, the acceptance and efficiency depend on both $y$ and \pt.  We chose the
$y$ interval such that our efficiency and acceptance depend only on \pt.  Candidates are selected by
placing restrictions on the topology of the decay.   These have been optimized to obtain maximum
mass signal significance and are listed in Table \ref{tCuts}.

\begin{figure}[!bt]
\centering
\includegraphics[width=\textwidth]{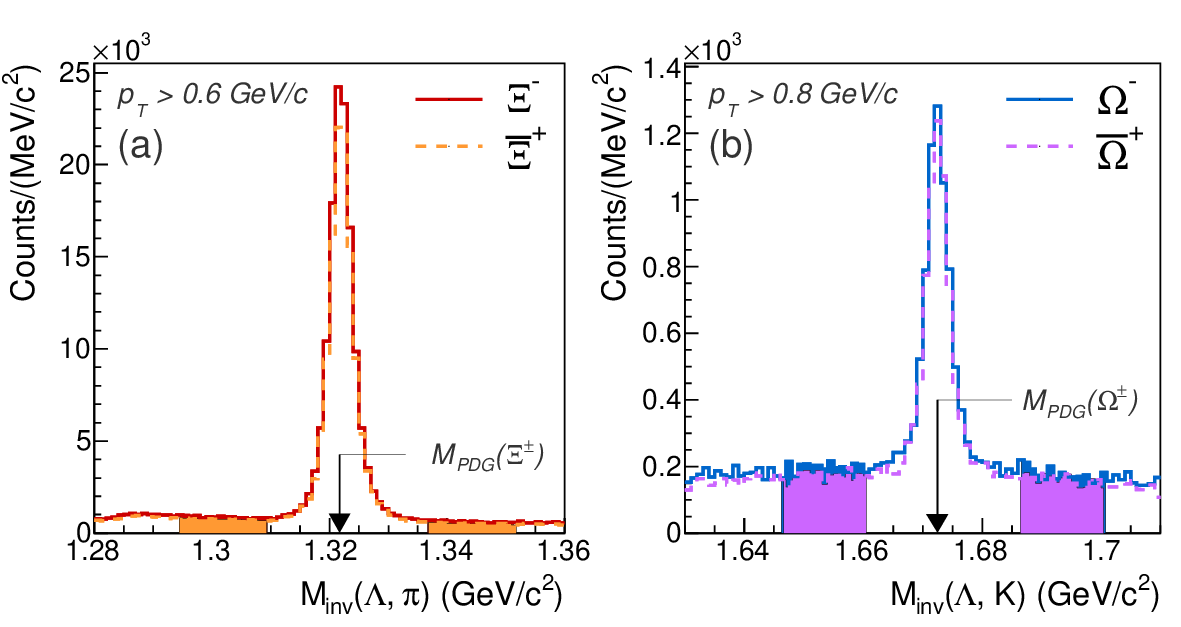}
\caption{\label{fMassPeaks} The invariant mass distributions of \X\ (a) and \Om\ (b) baryon
candidates (solid histograms) and their antiparticles (dashed histograms).  Also marked (in shaded blocks) background sampling regions used in the signal extraction. The entire measured \pt\
range is presented. }
\end{figure}

The resulting invariant mass distributions for both species hypotheses are shown in Fig.~\ref{fMassPeaks}.  
The signal extraction method is described in detail in \cite{Aamodt:2011zza}.  
The signal is extracted using a bin-counting method and then corrected for detector efficiency and 
acceptance using PYTHIA Perugia 0 \cite{Skands:2010ak} generated Monte Carlo events propagated through 
ALICE using GEANT3 \cite{Brun:1994aa}.  

\begin{table}[!tb]
\caption{Selection criteria parameters for V0 ($\Lambda$) and cascades (\Xis\ and \Oms) presented in
this Letter. If a criterion for \Xis\ and \Oms\ finding differs, the criterion for \Oms\ hypothesis
is in parentheses.
DCA stands for ``distance of closest approach," and PV for ``primary event vertex." The fiducial volume is defined by the coordinate of the decay vertex position, the transverse radius, $R_{2D}$.
$\theta$ is the angle between the momentum vector of the reconstructed V0 or cascade, and the line
segment bound by the decay and primary vertices. 
For cascades, the curvature of the particle's trajectory is neglected.
\label{tCuts}
}
~\newline
\centering
\begin{tabular}{ll}
\hline
\noalign{\smallskip}
 \multicolumn{2}{l}{\bf ~~~V0 finding criteria}                      \\
\noalign{\smallskip}
\hline
\noalign{\smallskip}
 \it DCA (h$^\pm$ to PV)                        & $>0.04$ (0.03) cm                           \\ 
 \it DCA (h$^-$ to h$^+$)                       & $<1.6$ standard deviations                  \\
 \it $\Lambda$ mass ($m_{\mathrm{V0}}$)         & $1.110<m_{\mathrm{V0}}<$ 1.122 \GeVmass     \\
 \it Fiducial volume (R$_{2D}$)                 & 1.4 $<$ $R_{2D}<$ 100 cm                    \\
 \it V0 pointing angle                          & $\cos\theta_{V0}>$ 0.97                     \\

\noalign{\smallskip}
\hline
\noalign{\smallskip}
\multicolumn{2}{l}{\bf ~~~Cascade finding criteria}                  \\
\hline
\noalign{\smallskip} 
  \it DCA ($\pi^{\pm}$ (K$^{\pm}$) to PV)       & $>0.05$ cm\\
  \it DCA (V0 to PV)                            & $>0.07$ cm\\
  \it DCA ($\pi^{\pm}$ (K$^{\pm}$) to V0)       & $< 1.6~(1.0)$ cm\\    
  \it Fiducial volume (R$_{2D}$)                & 0.8 (0.6) $<$ $R_{2D}<$ 100 cm\\
  \it Cascade pointing angle                    & $\cos\theta_{casc}>$ 0.97\\
\noalign{\smallskip}                     
\hline 
\end{tabular}
\end{table}

\section{Systematic uncertainties}
\label{sec:SystUncertainties}
There are two types of systematic uncertainties in the resulting particle spectra: \pt-dependent 
systematic uncertainties that are due to the efficiency determination and the signal quality at a 
given \pt, and the \pt-independent uncertainties due to normalization and other factors explained below.  

The point-to-point systematic uncertainties vary between 1-4~\%  for \Xis, and 1-9~\% for \Oms, with minimum uncertainty found at  
\mbox{\pt~=~1.5-4.0~\GeVc} for both species.  The \pt-independent systematic uncertainties stem from several sources and reflect the following: 
\begin{itemize}
	\item the uncertainty in determination of the material thickness traversed by the particles (material budget), 4~\%;
	\item the use of FLUKA \cite{Battistoni:2007zzb}\cite{Ferrari:2005zk} to correct \cite{Aamodt:2010dx} the antiproton 
		absorption cross section in GEANT3 \cite{Brun:1994aa}, 1~\%; 
	\item the uncertainty in TPC particle identification via energy loss, 1.5~\%; 
	\item the uncertainty on the track selection in the TPC, through the restriction on the number of TPC pad plane 
		clusters used in particle reconstruction, 3~\%;
	\item in the case of \Oms, the removal of cascades that fit the \Xis\ baryon hypothesis, 1~\%.
\end{itemize}

The limited \pt-coverage and determination of the total number of inelastic events used for yield normalization lead to an additional uncertainty in the particle yields and mean \pt\ (\meanpt) values.  
The INEL normalization \cite{Alice:2012xsec} leads to a +7.0~\% and --3.5~\% uncertainty on the yield 
for all measured particles, while the limited \pt\ coverage causes a 4.5~\% uncertainty on the \meanpt\ of all species, 5.5~\% uncertainty on the yield of \Xis\ baryons, and 6.5~\% on the yield of \Oms.  
While the systematic uncertainties (both \pt-dependent and \pt-independent) associated with each spectrum point 
affect the determination of \meanpt, the systematic uncertainty 
on the \meanpt\ for all species is dominated by the 4.5~\% error due to the limited \pt\ coverage.
Similarly, the systematic uncertainty on the yields is dominated by the uncertainties due to low-\pt\ 
extrapolation and event normalization.

\section{Results}
\label{sec:Results}

\subsection{Corrected \pt\ spectra and Tsallis fits}
\label{ssec:CorrectedSpectra}

\begin{figure}[!bt]
\centering
\includegraphics[width=3.50in]{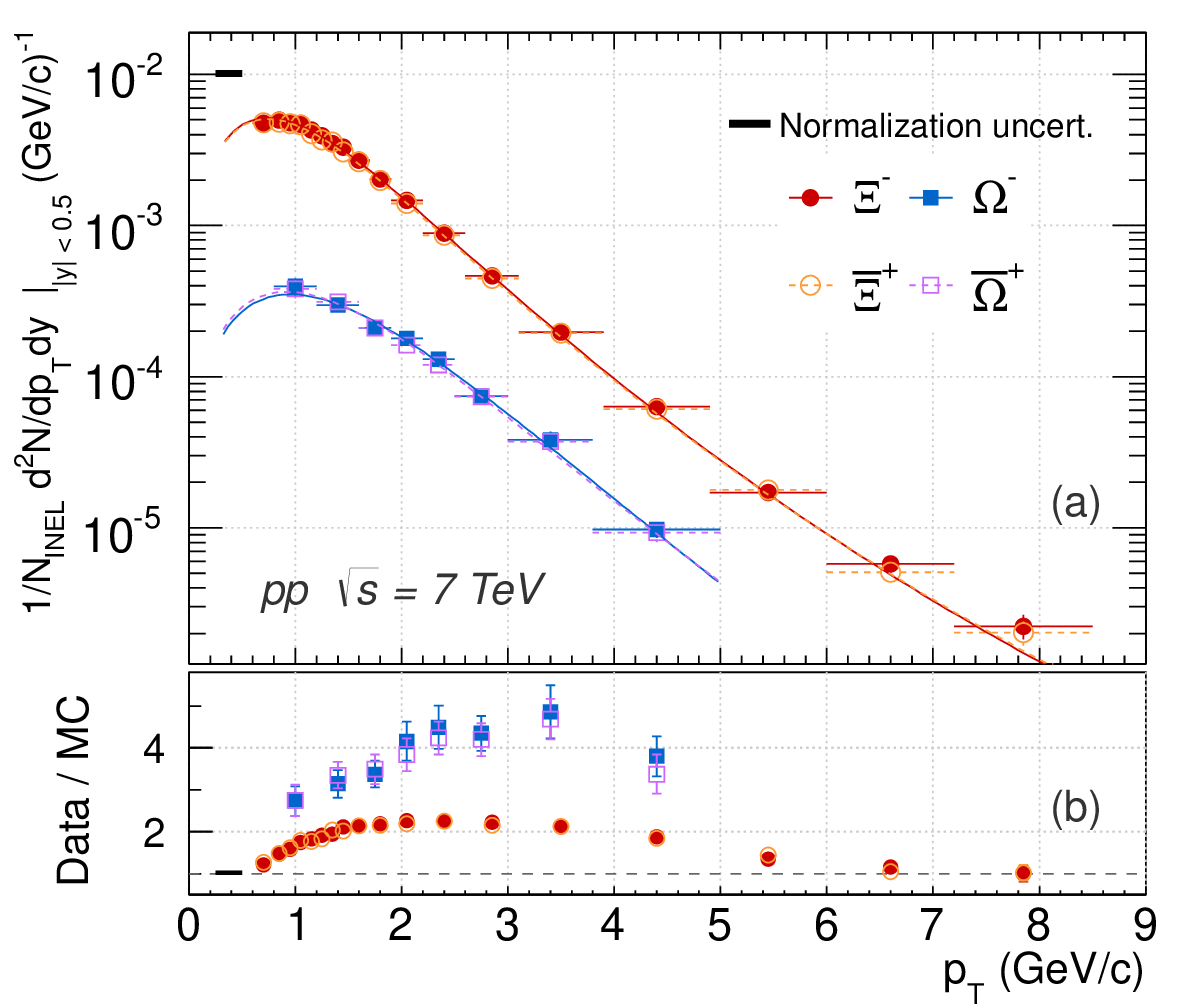}
\caption{\label{fSpectra} (a) \X\ and \Om\ baryon (solid circles and squares, respectively) 
and their antiparticle (open symbols) spectra, shown with Tsallis fits.  
(b) Experimental data to Monte Carlo (PYTHIA Perugia 2011) comparison.  The errors are added in quadrature.  The normalization uncertainty is shown as a black band.}
\end{figure}

The corrected multi-strange baryon yields per \pt\ bin per unit rapidity 
(1/N$_{\mathrm{INEL}}\times\mathrm{d}^2N/\mathrm{d}y\mathrm{d}$\pt) 
are shown in Fig.~\ref{fSpectra}(a).  
They span from \pt~=~0.6 to \mbox{\pt~=~8.5 \GeVc} in the case of \X\ and \Ix\ baryons 
and from \pt~=~0.8 to \mbox{\pt~=~5~\GeVc} for \Om\ and \Mo\ baryons.   
The Tsallis function is used for fitting the spectra, as the measured \pt\ range covers both 
soft-physics and fragmentation particle production regions. The functional form is shown 
below:

$$\frac{\mathrm{d}^2N}{\mathrm{d}y\mathrm{d} p_{\mathrm{T}}}=\frac{(n-1)(n-2)}{nT[nT+m_\mathrm{0}(n-2)]}\times
\frac{\mathrm{d}N}{\mathrm{d}y}\times p_{\mathrm{T}}\times\left(1+\frac{m_{\mathrm{T}}-m_{\mathrm{0}}}{nT}\right)^{-n}$$

where $T$, $n$, and \dndy\ (\dndy\ representing the particle yield per unit rapidity) are fit
parameters,  \mbox{$m_{\mathrm{T}}=\sqrt{m_0^2+p_{\mathrm{T}}^2}$}, and $m_0$ denotes the particle mass.  

The function is grounded in Tsallis statistics \cite{Tsallis:1987eu}; it approximates an exponential 
component (represented by the $T$ parameter), as well as a power-law dependence for the high-\pt\ tail.  
In Table~\ref{table_pars}, we list the fit results for each particle and antiparticle and the 
corresponding extrapolated \dndy\ and \meanpt{}.  

The central values of the fit parameters, listed in Table~\ref{table_pars}, are obtained using the 
statistical error only.  
The low-\pt\ extrapolation of the yield from the Tsallis fit is $\sim23~\%$ for \Xis\ and $\sim26~\%$ for \Oms. 
The value of \meanpt\ for each particle was computed using the fit over the entire \pt\ range 
including the extrapolation.
The antiparticle to particle ratios were found to be compatible with unity at all \pt.  

\begin{table*}[!bt]
\caption{Tsallis fit parameters, yields, and \meanpt\ for each particle species at $|y|<0.5$, as well 
as yields and \meanpt\ extracted from PYTHIA Perugia 2011 \cite{Skands:2010ak} simulations.  
Statistical and then systematic uncertainties for the experimental values are listed.  
Sufficient statistics were generated to have less than $1~\%$ error on all model values.}
\label{table_pars}
~\newline
\centering
\resizebox{1.00\textwidth}{!}{ 
\begin{tabular}{lccccccc}
\hline
\noalign{\smallskip}
Particle & $T$ (MeV)   & $n$ & $\chi^2$/NDF & \dndy\ $\times10^{3}$  & \meanpt{} (\GeVc) & \dndy\ $\times10^{3}$  & \meanpt{} (\GeVc) \\
         &             &     &              &  data                  &   data            &  P2011                 & P2011 \\
\noalign{\smallskip}
\hline
\noalign{\smallskip}
\X\     & $344\pm5\pm10$       & $10.8\pm0.4\pm0.8$      & 17.4/15 & $8.0\pm0.1^{+0.7}_{-0.5}$      & $1.21\pm0.01\pm0.06$ &$5.38$  &$1.02$\\
\noalign{\smallskip}
\Ix\    & $339\pm5\pm9$        & $10.4\pm0.4\pm0.5$      & 14.4/15 & $7.8\pm0.1^{+0.7}_{-0.5}$      & $1.21\pm0.01\pm0.06$ &$5.21$  &$1.02$\\
\noalign{\bigskip}
\Om\    & $460\pm40\pm60$      & $20\pm9\pm8$            & 8.8/5   & $0.67\pm0.03^{+0.08}_{-0.07}$  & $1.47\pm0.03\pm0.09$ &$0.276$ &$1.14$\\
\noalign{\smallskip}
\Mo\    & $430\pm30\pm40$      & $14\pm5\pm6$            & 7.0/5   & $0.68\pm0.03^{+0.08}_{-0.06}$  & $1.44\pm0.03\pm0.08$ &$0.266$ &$1.16$\\
\noalign{\smallskip}                     
\hline 
\end{tabular}
}
\end{table*}

\subsection{Excitation functions}
\label{ssec:ExcitationFunctions}

\begin{figure}[!tb]
\centering
\includegraphics[width=3.50in]{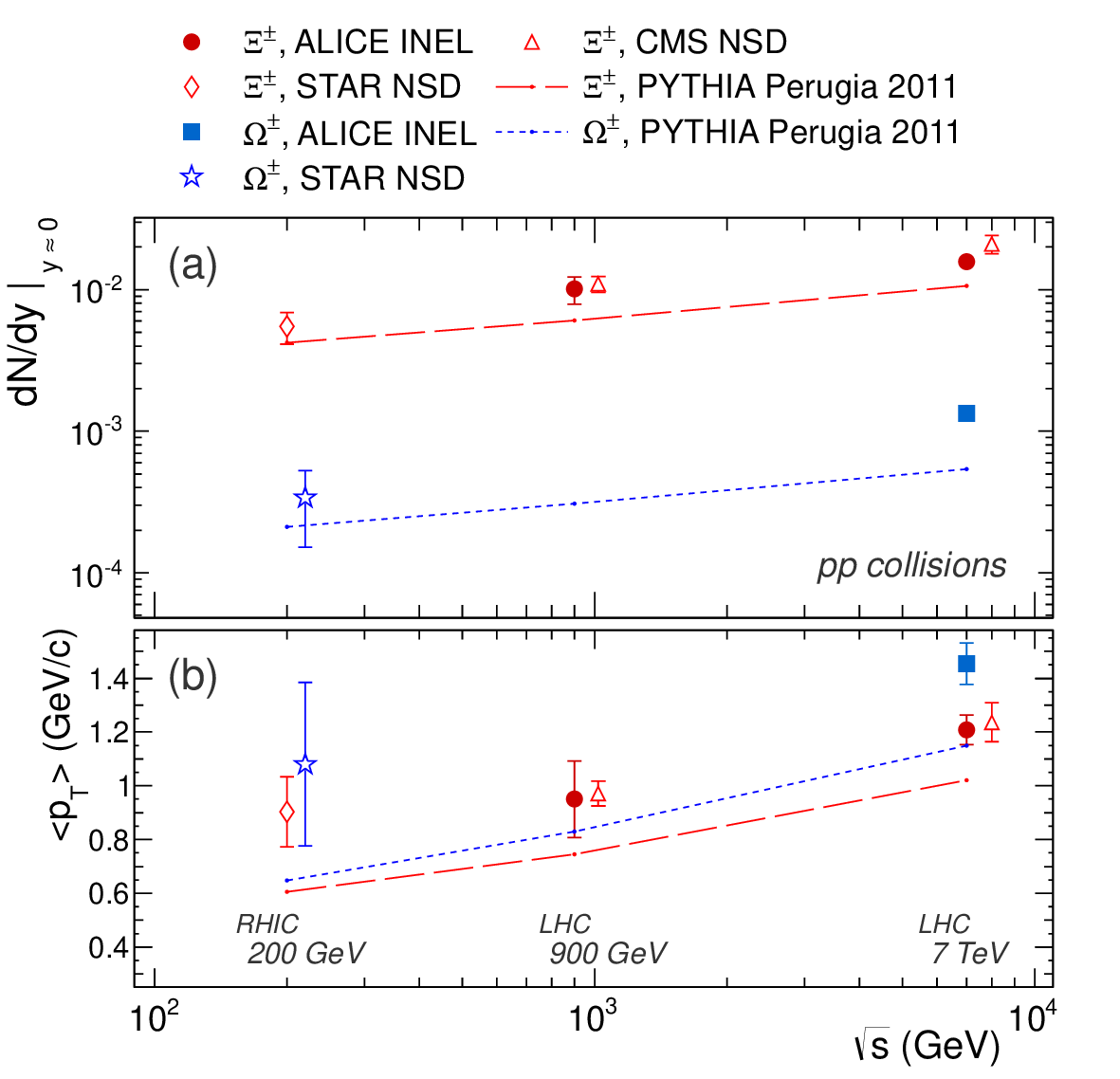}
\caption{\label{fdNdy_sqrts} (a)  \dndy\ and (b) \meanpt\ of \Xis\ and \Oms\ as a function of collision energy. 
The STAR and CMS data are normalized to NSD (see text) events, STAR \Xis\ and \Oms\ are represented 
by open rhombuses and stars, respectively.  CMS \Xis\ measurements are shown as open triangles, 
and ALICE \Xis\ and \Oms\ as filled circles and squares.  
Multi-strange baryons produced using PYTHIA Perugia 2011 simulation (\Xis\ baryons as a long-dashed curve 
and \Oms\ baryons as a dashed curve) are plotted for reference. 
The uncertainties are added in quadrature.}
\end{figure}

Our measurements of multi-strange baryons can be placed within the broader context of existing \pp\
collision data. We compare to multi-strange baryon yields in \pp\ collisions measured by the STAR
Collaboration at $\sqrt{s}$ = 0.2~TeV \cite{Abelev:2006c},  and also to the data obtained by ALICE and CMS at \nineH\ 
\cite{Aamodt:2011zza}\cite{Khachatryan:2011tm}.   There are also data from p$\mathrm{\bar{p}}$ collisions, obtained by the CDF \cite{Aaltonen:2011wz} and UA5 \cite{Ansorge:1989ba} Collaborations. We omit the comparison to these data due to a significantly different kinematic range of the experiments.
For STAR, ALICE, and CMS data, an increase in \dndy\  as a function of collision energy is observed, presented in 
Fig.~\ref{fdNdy_sqrts}(a).  We note that the CMS Collaboration used non-single-diffractive events (NSD)
 to normalize the yield, while in ALICE a normalization to the inelastic events (INEL) was used. 
For a direct comparison at LHC energies, the INEL \Xis\ yield has to be scaled up by 26~\% to get
the yield normalized to NSD events \cite{Alice:2012xsec}. 
After scaling, the \Xis\ yields per unit of rapidity obtained by ALICE agree with those published by
CMS \cite{Khachatryan:2011tm}.  
For \X\ baryons and antibaryons, we also observe a slight rise in mean \pt\ with collision energy,
as seen in Fig. \ref{fdNdy_sqrts}(b).   
The \meanpt\ of \Oms\ baryons at \seven\ is consistent with 0.2 TeV data, where \Oms\ and \Xis\ \meanpt\ 
were consistent within large experimental error.  
Due to the precision of the current measurements, a significant separation between the \meanpt\
of \Oms\ and \Xis\ is observed in \seven\ \pp\ collisions. 

\subsection{(\Om+\Mo) / (\X+\Ix) ratio}
\label{ssec:RatioOmegaToXi}

\begin{figure}[!bt]
\centering
\includegraphics[width=3.50in]{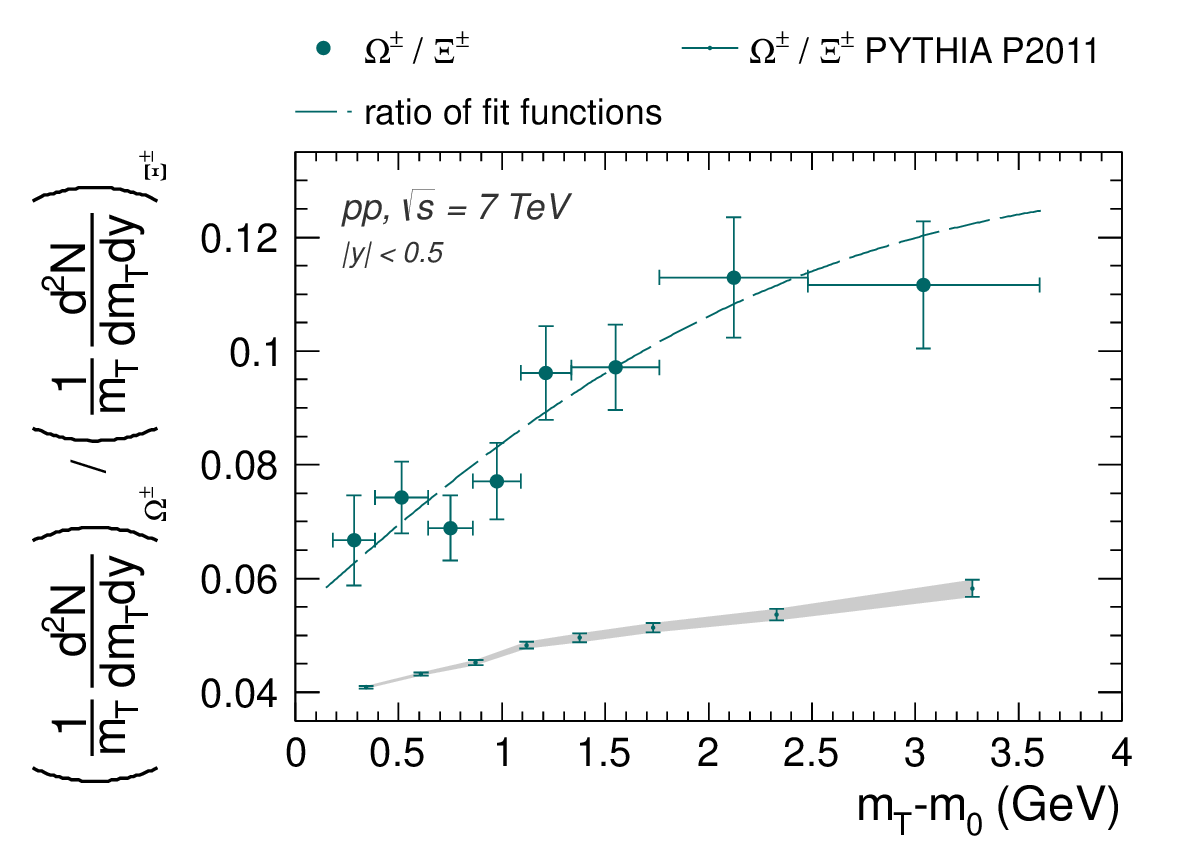}
\caption{\label{fOmegaOverXi} (\Om+\Mo) to (\X+\Ix) ratio in \seven\ \pp\ events as a function of (\mTmZero).  
Experimental data (closed symbols: data points; dashed curve: ratio of Tsallis fits), 
and PYTHIA Perugia 2011 simulation (solid curve).  Errors on experimental points were added in quadrature.}
\end{figure}

The composition of \X\ and \Om\ baryons differs only by one valence quark flavour:
the $d$-quark in \X\ is replaced by the $s$-quark in \Om.  
To investigate possible differences in the production mechanism of multi-strange baryons with and 
without the non-strange quark, we study the ratio of (\Om+\Mo) to (\X+\Ix) baryons as a function of \pt.  
The dependence on particle mass is reduced by constructing spectra as a function of (\mTmZero)
for each baryon species.  To increase the statistical significance of the measurement, for this ratio, the particle and antiparticle spectra are combined.
The ratio of the combined spectra,  (\Om+\Mo) / (\X+\Ix), is shown in Fig. \ref{fOmegaOverXi}.  
We observe an increase in the ratio up to $(\mTmZero) \sim 1.5$~GeV (which corresponds roughly to \pt\ of 
3 \GeVc\ for either of the baryons), with a possible slope change at a higher (\mTmZero).  
The ratio was composed to investigate the possible saturation of the $s$-quark production with respect to production of the non-strange quarks, which would be indicated by the flattening of the (\Om+\Mo) / (\X+\Ix) ratio.  We found that the currently available data does not allow for a firm conclusion. 

\subsection{Comparisons to PYTHIA Perugia 2011}
\label{ssec:ComparisonsPYTHIA}

The production of strangeness in \pp\ collisions is not well described by the currently available models.  
In particular, we compare the obtained data to particle spectra from PYTHIA \cite{Sjostrand:2006za}, 
an event generator based on the leading order (LO) perturbative Quantum Chromo-Dynamics (pQCD).   
PYTHIA is available in different tunes, for example those listed in \cite{Skands:2010ak}, each reflecting a 
distinct aspect of particle production inferred from experimental data.  Several tunes were tested, among them PYTHIA Z1, Z2 \cite{Field:2011iq}, and Perugia 0 \cite{Skands:2010ak} tunes.  
These tunes were several times to an order of magnitude below the measured multi-strange spectra and yields (up to a factor 4 for \Xis, 15 for \Oms) and thus were abandoned. 

In this Letter, we use the central PYTHIA Perugia 2011 (P2011), one of the more recent PYTHIA 6.4 tunes, 
which describes the 7 TeV \pp\ charged particle spectra reasonably well.  
P2011 is tuned to the multiplicity and charged particle \pt\ distributions from 2010 LHC data (as were other tunes tested), utilizes the CTEQ5L parton distribution function, and differs 
from other PYTHIA tunes by a significant increase in multi-strange baryon yields. 
This is achieved mainly by removing the baryon suppression inherently present in the built-in ``pop corn" 
meson creation mechanism, but also by tuning the relative $u$ and $d$ vs. $s$-quark production rates, 
and adjusting the suppression of the diquark-antiquark hadron production scale \cite{Skands:2010ak}.  
The ``pop corn" mechanism, used to describe the hadron production in $\mathrm{e}^+\mathrm{e}^-$ collisions 
via chromoelectric flux tubes \cite{Casher:1978wy}, suppresses baryon production
by favouring soft quark-antiquark pairing into mesons \cite{Chun:1992qs}. 
The mechanism was removed in order to approximate more closely the multi-strange yields from LEP
experiments \cite{Skands:2010ak}.   

Although the charged-particle multiplicities are reasonably described, as mentioned above, PYTHIA tunes tend to be several times to an order of magnitude below measured multi-strange values.  P2011 significantly   underestimates multi-strange particle yields, as seen in Table~\ref{table_pars}, and does not reproduce the spectral shapes of either  
\X\ or \Om\ baryons, with two exceptions.  The model describes the high \pt\ tail of the \Xis\ distribution and approaches the \Xis\ distribution below \pt\ $<0.85$ \GeVc,
as shown in Fig. \ref{fSpectra}(b). 

P2011 also underpredicts \meanpt\ of multi-strange baryons at all energies (Fig. \ref{fdNdy_sqrts}(b)), and incorrectly models 
the increase in \dndy\  as a function of centre-of-mass energy (Fig. \ref{fdNdy_sqrts}(a)). 
Indeed, in experimental data \dndy\ increases by nearly a factor of three from \twoH\ collisions to 
those at \seven\ (factor 35 increase in energy), while P2011 predicts a more modest gain.  
In both experimental data and P2011, a power-law increase of \X\ baryon yield is seen as a function of $\sqrt{s}$. Moreover, P2011 does not reproduce the relative \Oms/\Xis\ spectral shape, nor the absolute value, although 
the ratio does increase with increased \pt, as shown in Fig. \ref{fOmegaOverXi}.

\section{Conclusions}
\label{sec:Summary}

Our precise measurements of \X, \Ix, \Om, and \Mo in \seven\ \pp\ collisions are a benchmark for 
improving future modelling efforts, including valuable checks on possible hadron production mechanisms, 
such as the flux-tube mechanism.  In addition, the \pt\ reach of the data to model comparison is the highest ever achieved for multi-strange baryons. 
The relative production of doubly-strange vs. triply-strange baryons introduces a further constraint
on the \pt\ dependence of particle production from flavour-differentiated quarks. These considerations 
may enable a better insight into \pp\ collision dynamics, which in turn will serve as a reference 
for better understanding of fundamental interactions underlying particle creation mechanisms in \pp\ 
collisions.
%
\newenvironment{acknowledgement}{\relax}{\relax}
\begin{acknowledgement}
\section*{Acknowledgements}
The ALICE Collaboration would like to thank all its engineers and technicians for their invaluable contributions to the construction of the experiment and the CERN accelerator teams for the outstanding performance of the LHC complex.
\\
The ALICE Collaboration acknowledges the following funding agencies for their support in building and
running the ALICE detector:
 \\
Calouste Gulbenkian Foundation from Lisbon and Swiss Fonds Kidagan, Armenia;
 \\
Conselho Nacional de Desenvolvimento Cient\'{\i}fico e Tecnol\'{o}gico (CNPq), Financiadora de Estudos e Projetos (FINEP),
Funda\c{c}\~{a}o de Amparo \`{a} Pesquisa do Estado de S\~{a}o Paulo (FAPESP);
 \\
National Natural Science Foundation of China (NSFC), the Chinese Ministry of Education (CMOE)
and the Ministry of Science and Technology of China (MSTC);
 \\
Ministry of Education and Youth of the Czech Republic;
 \\
Danish Natural Science Research Council, the Carlsberg Foundation and the Danish National Research Foundation;
 \\
The European Research Council under the European Community's Seventh Framework Programme;
 \\
Helsinki Institute of Physics and the Academy of Finland;
 \\
French CNRS-IN2P3, the `Region Pays de Loire', `Region Alsace', `Region Auvergne' and CEA, France;
 \\
German BMBF and the Helmholtz Association;
\\
General Secretariat for Research and Technology, Ministry of
Development, Greece;
\\
Hungarian OTKA and National Office for Research and Technology (NKTH);
 \\
Department of Atomic Energy and Department of Science and Technology of the Government of India;
 \\
Istituto Nazionale di Fisica Nucleare (INFN) of Italy;
 \\
MEXT Grant-in-Aid for Specially Promoted Research, Ja\-pan;
 \\
Joint Institute for Nuclear Research, Dubna;
 \\
National Research Foundation of Korea (NRF);
 \\
CONACYT, DGAPA, M\'{e}xico, ALFA-EC and the HELEN Program (High-Energy physics Latin-American--European Network);
 \\
Stichting voor Fundamenteel Onderzoek der Materie (FOM) and the Nederlandse Organisatie voor Wetenschappelijk Onderzoek (NWO), Netherlands;
 \\
Research Council of Norway (NFR);
 \\
Polish Ministry of Science and Higher Education;
 \\
National Authority for Scientific Research - NASR (Autoritatea Na\c{t}ional\u{a} pentru Cercetare \c{S}tiin\c{t}ific\u{a} - ANCS);
 \\
Federal Agency of Science of the Ministry of Education and Science of Russian Federation, International Science and
Technology Center, Russian Academy of Sciences, Russian Federal Agency of Atomic Energy, Russian Federal Agency for Science and Innovations and CERN-INTAS;
 \\
Ministry of Education of Slovakia;
 \\
Department of Science and Technology, South Africa;
 \\
CIEMAT, EELA, Ministerio de Educaci\'{o}n y Ciencia of Spain, Xunta de Galicia (Conseller\'{\i}a de Educaci\'{o}n),
CEA\-DEN, Cubaenerg\'{\i}a, Cuba, and IAEA (International Atomic Energy Agency);
 \\
Swedish Research Council (VR) and Knut $\&$ Alice Wallenberg
Foundation (KAW);
 \\
Ukraine Ministry of Education and Science;
 \\
United Kingdom Science and Technology Facilities Council (STFC);
 \\
The United States Department of Energy, the United States National
Science Foundation, the State of Texas, and the State of Ohio.
\end{acknowledgement}
%

\bibliographystyle{utphys}
\bibliography{alice_mspp_cds}

\newpage
\appendix
\section{The ALICE Collaboration}
\label{app:collab}

\begingroup
\small
\begin{flushleft}
B.~Abelev\Irefn{org1234}\And
J.~Adam\Irefn{org1274}\And
D.~Adamov\'{a}\Irefn{org1283}\And
A.M.~Adare\Irefn{org1260}\And
M.M.~Aggarwal\Irefn{org1157}\And
G.~Aglieri~Rinella\Irefn{org1192}\And
A.G.~Agocs\Irefn{org1143}\And
A.~Agostinelli\Irefn{org1132}\And
S.~Aguilar~Salazar\Irefn{org1247}\And
Z.~Ahammed\Irefn{org1225}
% ...
\renewcommand\labelenumi{\textsuperscript{\theenumi}~}
\section*{Affiliation notes}
\renewcommand\theenumi{\roman{enumi}}
\begin{Authlist}
\item \Adef{M.V.Lomonosov Moscow State University, D.V.Skobeltsyn Institute of Nuclear Physics, Moscow, Russia}Also at: M.V.Lomonosov Moscow State University, D.V.Skobeltsyn Institute of Nuclear Physics, Moscow, Russia
\item \Adef{Institute of Nuclear Sciences, Belgrade, Serbia}Also at: "Vin\v{c}a" Institute of Nuclear Sciences, Belgrade, Serbia
\end{Authlist}
\section*{Collaboration Institutes}
\renewcommand\theenumi{\arabic{enumi}~}
\begin{Authlist}
\item \Idef{org1279}Benem\'{e}rita Universidad Aut\'{o}noma de Puebla, Puebla, Mexico
\item \Idef{org1220}Bogolyubov Institute for Theoretical Physics, Kiev, Ukraine
\item \Idef{org1262}Budker Institute for Nuclear Physics, Novosibirsk, Russia
\item \Idef{org1292}California Polytechnic State University, San Luis Obispo, California, United States
\item \Idef{org14939}Centre de Calcul de l'IN2P3, Villeurbanne, France
\item \Idef{org1197}Centro de Aplicaciones Tecnol\'{o}gicas y Desarrollo Nuclear (CEADEN), Havana, Cuba
\item \Idef{org1242}Centro de Investigaciones Energ\'{e}ticas Medioambientales y Tecnol\'{o}gicas (CIEMAT), Madrid, Spain
\item \Idef{org1244}Centro de Investigaci\'{o}n y de Estudios Avanzados (CINVESTAV), Mexico City and M\'{e}rida, Mexico
\item \Idef{org1335}Centro Fermi -- Centro Studi e Ricerche e Museo Storico della Fisica ``Enrico Fermi'', Rome, Italy
\item \Idef{org17347}Chicago State University, Chicago, United States
\item \Idef{org1288}Commissariat \`{a} l'Energie Atomique, IRFU, Saclay, France
\item \Idef{org1294}Departamento de F\'{\i}sica de Part\'{\i}culas and IGFAE, Universidad de Santiago de Compostela, Santiago de Compostela, Spain
\item \Idef{org1106}Department of Physics Aligarh Muslim University, Aligarh, India
\item \Idef{org1121}Department of Physics and Technology, University of Bergen, Bergen, Norway
\item \Idef{org1162}Department of Physics, Ohio State University, Columbus, Ohio, United States
\item \Idef{org1300}Department of Physics, Sejong University, Seoul, South Korea
\item \Idef{org1268}Department of Physics, University of Oslo, Oslo, Norway
\item \Idef{org1145}Dipartimento di Fisica dell'Universit\`{a} and Sezione INFN, Cagliari, Italy
\item \Idef{org1270}Dipartimento di Fisica dell'Universit\`{a} and Sezione INFN, Padova, Italy
\item \Idef{org1315}Dipartimento di Fisica dell'Universit\`{a} and Sezione INFN, Trieste, Italy
\item \Idef{org1132}Dipartimento di Fisica dell'Universit\`{a} and Sezione INFN, Bologna, Italy
\item \Idef{org1285}Dipartimento di Fisica dell'Universit\`{a} `La Sapienza' and Sezione INFN, Rome, Italy
\item \Idef{org1154}Dipartimento di Fisica e Astronomia dell'Universit\`{a} and Sezione INFN, Catania, Italy
\item \Idef{org1290}Dipartimento di Fisica `E.R.~Caianiello' dell'Universit\`{a} and Gruppo Collegato INFN, Salerno, Italy
\item \Idef{org1312}Dipartimento di Fisica Sperimentale dell'Universit\`{a} and Sezione INFN, Turin, Italy
\item \Idef{org1103}Dipartimento di Scienze e Tecnologie Avanzate dell'Universit\`{a} del Piemonte Orientale and Gruppo Collegato INFN, Alessandria, Italy
\item \Idef{org1114}Dipartimento Interateneo di Fisica `M.~Merlin' and Sezione INFN, Bari, Italy
\item \Idef{org1237}Division of Experimental High Energy Physics, University of Lund, Lund, Sweden
\item \Idef{org1192}European Organization for Nuclear Research (CERN), Geneva, Switzerland
\item \Idef{org1227}Fachhochschule K\"{o}ln, K\"{o}ln, Germany
\item \Idef{org1122}Faculty of Engineering, Bergen University College, Bergen, Norway
\item \Idef{org1136}Faculty of Mathematics, Physics and Informatics, Comenius University, Bratislava, Slovakia
\item \Idef{org1274}Faculty of Nuclear Sciences and Physical Engineering, Czech Technical University in Prague, Prague, Czech Republic
\item \Idef{org1229}Faculty of Science, P.J.~\v{S}af\'{a}rik University, Ko\v{s}ice, Slovakia
\item \Idef{org1184}Frankfurt Institute for Advanced Studies, Johann Wolfgang Goethe-Universit\"{a}t Frankfurt, Frankfurt, Germany
\item \Idef{org1215}Gangneung-Wonju National University, Gangneung, South Korea
\item \Idef{org1212}Helsinki Institute of Physics (HIP) and University of Jyv\"{a}skyl\"{a}, Jyv\"{a}skyl\"{a}, Finland
\item \Idef{org1203}Hiroshima University, Hiroshima, Japan
\item \Idef{org1329}Hua-Zhong Normal University, Wuhan, China
\item \Idef{org1254}Indian Institute of Technology, Mumbai, India
\item \Idef{org36378}Indian Institute of Technology Indore (IIT), Indore, India
\item \Idef{org1266}Institut de Physique Nucl\'{e}aire d'Orsay (IPNO), Universit\'{e} Paris-Sud, CNRS-IN2P3, Orsay, France
\item \Idef{org1277}Institute for High Energy Physics, Protvino, Russia
\item \Idef{org1249}Institute for Nuclear Research, Academy of Sciences, Moscow, Russia
\item \Idef{org1320}Nikhef, National Institute for Subatomic Physics and Institute for Subatomic Physics of Utrecht University, Utrecht, Netherlands
\item \Idef{org1250}Institute for Theoretical and Experimental Physics, Moscow, Russia
\item \Idef{org1230}Institute of Experimental Physics, Slovak Academy of Sciences, Ko\v{s}ice, Slovakia
\item \Idef{org1127}Institute of Physics, Bhubaneswar, India
\item \Idef{org1275}Institute of Physics, Academy of Sciences of the Czech Republic, Prague, Czech Republic
\item \Idef{org1139}Institute of Space Sciences (ISS), Bucharest, Romania
\item \Idef{org27399}Institut f\"{u}r Informatik, Johann Wolfgang Goethe-Universit\"{a}t Frankfurt, Frankfurt, Germany
\item \Idef{org1185}Institut f\"{u}r Kernphysik, Johann Wolfgang Goethe-Universit\"{a}t Frankfurt, Frankfurt, Germany
\item \Idef{org1177}Institut f\"{u}r Kernphysik, Technische Universit\"{a}t Darmstadt, Darmstadt, Germany
\item \Idef{org1256}Institut f\"{u}r Kernphysik, Westf\"{a}lische Wilhelms-Universit\"{a}t M\"{u}nster, M\"{u}nster, Germany
\item \Idef{org1246}Instituto de Ciencias Nucleares, Universidad Nacional Aut\'{o}noma de M\'{e}xico, Mexico City, Mexico
\item \Idef{org1247}Instituto de F\'{\i}sica, Universidad Nacional Aut\'{o}noma de M\'{e}xico, Mexico City, Mexico
\item \Idef{org23333}Institut of Theoretical Physics, University of Wroclaw
\item \Idef{org1308}Institut Pluridisciplinaire Hubert Curien (IPHC), Universit\'{e} de Strasbourg, CNRS-IN2P3, Strasbourg, France
\item \Idef{org1182}Joint Institute for Nuclear Research (JINR), Dubna, Russia
\item \Idef{org1143}KFKI Research Institute for Particle and Nuclear Physics, Hungarian Academy of Sciences, Budapest, Hungary
\item \Idef{org1199}Kirchhoff-Institut f\"{u}r Physik, Ruprecht-Karls-Universit\"{a}t Heidelberg, Heidelberg, Germany
\item \Idef{org20954}Korea Institute of Science and Technology Information, Daejeon, South Korea
\item \Idef{org1160}Laboratoire de Physique Corpusculaire (LPC), Clermont Universit\'{e}, Universit\'{e} Blaise Pascal, CNRS--IN2P3, Clermont-Ferrand, France
\item \Idef{org1194}Laboratoire de Physique Subatomique et de Cosmologie (LPSC), Universit\'{e} Joseph Fourier, CNRS-IN2P3, Institut Polytechnique de Grenoble, Grenoble, France
\item \Idef{org1187}Laboratori Nazionali di Frascati, INFN, Frascati, Italy
\item \Idef{org1232}Laboratori Nazionali di Legnaro, INFN, Legnaro, Italy
\item \Idef{org1125}Lawrence Berkeley National Laboratory, Berkeley, California, United States
\item \Idef{org1234}Lawrence Livermore National Laboratory, Livermore, California, United States
\item \Idef{org1251}Moscow Engineering Physics Institute, Moscow, Russia
\item \Idef{org1140}National Institute for Physics and Nuclear Engineering, Bucharest, Romania
\item \Idef{org1165}Niels Bohr Institute, University of Copenhagen, Copenhagen, Denmark
\item \Idef{org1109}Nikhef, National Institute for Subatomic Physics, Amsterdam, Netherlands
\item \Idef{org1283}Nuclear Physics Institute, Academy of Sciences of the Czech Republic, \v{R}e\v{z} u Prahy, Czech Republic
\item \Idef{org1264}Oak Ridge National Laboratory, Oak Ridge, Tennessee, United States
\item \Idef{org1189}Petersburg Nuclear Physics Institute, Gatchina, Russia
\item \Idef{org1170}Physics Department, Creighton University, Omaha, Nebraska, United States
\item \Idef{org1157}Physics Department, Panjab University, Chandigarh, India
\item \Idef{org1112}Physics Department, University of Athens, Athens, Greece
\item \Idef{org1152}Physics Department, University of Cape Town, iThemba LABS, Cape Town, South Africa
\item \Idef{org1209}Physics Department, University of Jammu, Jammu, India
\item \Idef{org1207}Physics Department, University of Rajasthan, Jaipur, India
\item \Idef{org1200}Physikalisches Institut, Ruprecht-Karls-Universit\"{a}t Heidelberg, Heidelberg, Germany
\item \Idef{org1325}Purdue University, West Lafayette, Indiana, United States
\item \Idef{org1281}Pusan National University, Pusan, South Korea
\item \Idef{org1176}Research Division and ExtreMe Matter Institute EMMI, GSI Helmholtzzentrum f\"ur Schwerionenforschung, Darmstadt, Germany
\item \Idef{org1334}Rudjer Bo\v{s}kovi\'{c} Institute, Zagreb, Croatia
\item \Idef{org1298}Russian Federal Nuclear Center (VNIIEF), Sarov, Russia
\item \Idef{org1252}Russian Research Centre Kurchatov Institute, Moscow, Russia
\item \Idef{org1224}Saha Institute of Nuclear Physics, Kolkata, India
\item \Idef{org1130}School of Physics and Astronomy, University of Birmingham, Birmingham, United Kingdom
\item \Idef{org1338}Secci\'{o}n F\'{\i}sica, Departamento de Ciencias, Pontificia Universidad Cat\'{o}lica del Per\'{u}, Lima, Peru
\item \Idef{org1316}Sezione INFN, Trieste, Italy
\item \Idef{org1271}Sezione INFN, Padova, Italy
\item \Idef{org1313}Sezione INFN, Turin, Italy
\item \Idef{org1286}Sezione INFN, Rome, Italy
\item \Idef{org1146}Sezione INFN, Cagliari, Italy
\item \Idef{org1133}Sezione INFN, Bologna, Italy
\item \Idef{org1115}Sezione INFN, Bari, Italy
\item \Idef{org1155}Sezione INFN, Catania, Italy
\item \Idef{org1322}Soltan Institute for Nuclear Studies, Warsaw, Poland
\item \Idef{org36377}Nuclear Physics Group, STFC Daresbury Laboratory, Daresbury, United Kingdom
\item \Idef{org1258}SUBATECH, Ecole des Mines de Nantes, Universit\'{e} de Nantes, CNRS-IN2P3, Nantes, France
\item \Idef{org1304}Technical University of Split FESB, Split, Croatia
\item \Idef{org1168}The Henryk Niewodniczanski Institute of Nuclear Physics, Polish Academy of Sciences, Cracow, Poland
\item \Idef{org17361}The University of Texas at Austin, Physics Department, Austin, TX, United States
\item \Idef{org1173}Universidad Aut\'{o}noma de Sinaloa, Culiac\'{a}n, Mexico
\item \Idef{org1296}Universidade de S\~{a}o Paulo (USP), S\~{a}o Paulo, Brazil
\item \Idef{org1149}Universidade Estadual de Campinas (UNICAMP), Campinas, Brazil
\item \Idef{org1239}Universit\'{e} de Lyon, Universit\'{e} Lyon 1, CNRS/IN2P3, IPN-Lyon, Villeurbanne, France
\item \Idef{org1205}University of Houston, Houston, Texas, United States
\item \Idef{org20371}University of Technology and Austrian Academy of Sciences, Vienna, Austria
\item \Idef{org1222}University of Tennessee, Knoxville, Tennessee, United States
\item \Idef{org1310}University of Tokyo, Tokyo, Japan
\item \Idef{org1318}University of Tsukuba, Tsukuba, Japan
\item \Idef{org21360}Eberhard Karls Universit\"{a}t T\"{u}bingen, T\"{u}bingen, Germany
\item \Idef{org1225}Variable Energy Cyclotron Centre, Kolkata, India
\item \Idef{org1306}V.~Fock Institute for Physics, St. Petersburg State University, St. Petersburg, Russia
\item \Idef{org1323}Warsaw University of Technology, Warsaw, Poland
\item \Idef{org1179}Wayne State University, Detroit, Michigan, United States
\item \Idef{org1260}Yale University, New Haven, Connecticut, United States
\item \Idef{org1332}Yerevan Physics Institute, Yerevan, Armenia
\item \Idef{org15649}Yildiz Technical University, Istanbul, Turkey
\item \Idef{org1301}Yonsei University, Seoul, South Korea
\item \Idef{org1327}Zentrum f\"{u}r Technologietransfer und Telekommunikation (ZTT), Fachhochschule Worms, Worms, Germany
\end{Authlist}
\endgroup

%

\end{document}